\documentclass[a4paper]{jpconf}
\usepackage{amsmath}
\usepackage{graphicx}
\usepackage{wrapfig}
\usepackage{epsfig}

\begin{document}

\title{UNEDF: {\it Advanced Scientific Computing Collaboration Transforms the Low-Energy Nuclear Many-Body Problem}}

\author{H Nam$^{1}$,  M Stoitsov$^{1,2}$,  W Nazarewicz$^{1-3}$,
A Bulgac$^{4}$, G Hagen$^{1}$\\ M Kortelainen$^{1,2}$, P Maris$^{7}$, J C Pei$^{1,2}$, K J Roche$^{5}$, N Schunck$^{6}$, I Thompson$^{6}$,  J P Vary$^{7}$, and S M Wild$^{8}$
}

\address{
$^1$ Oak Ridge National Laboratory, P.O. Box 2008, Oak Ridge, TN 37831, USA \\
$^2$ Department of Physics and Astronomy, University of Tennessee Knoxville, TN 37996, USA \\
$^3$ Institute of Theoretical Physics, Warsaw University, ul. Ho\.{z}a 69, PL-00681, Warsaw, Poland\\
$^4$ Department of Physics, University of Washington, Seattle, WA 98195-1560, USA  \\
$^5$ Pacific Northwest National Laboratory, Richland, WA 99352, USA \\
$^6$ Lawrence Livermore National Laboratory, L-414, P.O. Box 808, Livermore, CA 94551, USA \\
$^7$ Department of Physics and Astronomy, Iowa State University, Ames, IA 50011-3160, USA \\
$^8$ Math \& Computer Science Division, Argonne National Laboratory, Argonne, IL 60439, USA \\
}

\ead{namha@ornl.gov}

\begin{abstract}
The demands of cutting-edge science are driving the need for larger and faster computing resources.  With the rapidly growing scale of computing systems and the prospect of technologically disruptive architectures to meet these needs, scientists face the challenge of effectively using complex computational resources to advance scientific discovery.  Multi-disciplinary collaborating networks of researchers with diverse scientific backgrounds are needed to address these complex challenges.  The UNEDF SciDAC collaboration of nuclear theorists, applied mathematicians, and computer scientists is developing a comprehensive description of nuclei and their reactions that delivers maximum predictive power with quantified uncertainties.  This paper describes UNEDF  and identifies attributes that classify it as a successful computational collaboration.  We illustrate significant milestones accomplished by UNEDF through integrative solutions using the most reliable theoretical approaches,  most advanced algorithms, and leadership-class computational resources.
\end{abstract}

\section{Introduction}
One of the discovery frontiers in  physics is to explain the nature of atomic nuclei. Apart from a plethora of basic science interests, this is also an essential component of energy, medical, and biological research, and national security. Nuclear physicists are working toward a fundamental and unified description of nuclei based on the underlying theory of the strong interactions, quantum chromodynamics, and to transform descriptive and highly phenomenological  models into predictive capability; in particular, allowing reliable extrapolations into regions that are not accessible by experiments. Ultimately, this would allow for accurate predictions of nuclear reactions, with significant impact on the development of advanced fission reactors and fusion energy sources, and in industrial and medical innovations through the use of stable isotopes and radioisotopes.

\begin{wrapfigure}[25]{l}{0.5\textwidth}
\begin{center}
\includegraphics[bb=0 0 420 420,clip,trim=.3cm .2cm 0cm 0cm,width=0.5\textwidth]{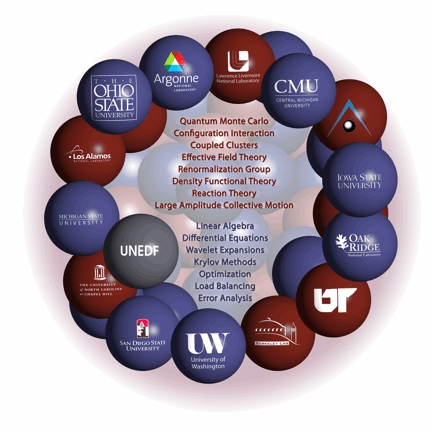}
\caption{UNEDF involves over 50 researchers from 9 universities and 7 national laboratories. Annually, it provides training to about 30 young researchers (postdocs and students).}
\label{collaboration}
\end{center}
\end{wrapfigure}

The UNEDF collaboration of nuclear theorists, applied mathematicians, and computer scientists (see Fig.~\ref{collaboration}) is making significant strides toward realizing this goal through a comprehensive study of all nuclei built on the latest advances in nuclear theory and scientific computing. UNEDF, which stands for ``Universal Nuclear Energy Density Functional,'' is a five-year SciDAC (``Scientific Discovery through Advanced Computing'') project \cite{UNEDF,NPN}.
The SciDAC program  (www.scidac.gov) has provided the opportunity for applied mathematicians and computer scientists to work collaboratively with physicists to develop and interconnect the most accurate knowledge of the strong nuclear interaction, high-precision theoretical approaches, scalable algorithms, and high-performance computing tools and libraries to enable scientific discoveries using leadership-class computing resources.  Working toward a predictive theory, the UNEDF project emphasizes the verification of methods and codes, the estimation of uncertainties, and the assessment of results.  An added and unexpected benefit of the UNEDF project has been the realization of new physics collaborations, identified through shared computational methods and needs.  Here we present an overview of  UNEDF, some significant milestones achieved through the UNEDF collaborative effort and the outlook for the future; more details and references can be found at the UNEDF website http://www.unedf.org.

\section{UNEDF Overview}
\begin{figure}[htdp]
\begin{minipage}{6in}
\begin{center}
\includegraphics[bb=0 0 350 300,clip,trim=0cm 0cm 0cm 0cm,width=0.85\textwidth]{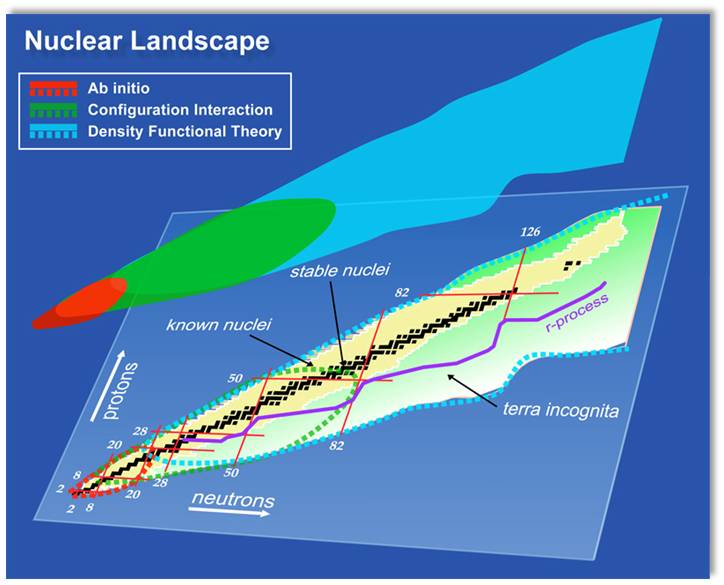}
\caption{Theoretical methods and computational techniques to solve the nuclear many-body problem across the nuclear landscape. The thick dotted lines indicate domains of major theoretical approaches to the nuclear many-body problem. For the lightest nuclei, ab initio calculations based on the bare nucleon-nucleon and three-nucleon interactions, are possible (red). Medium-mass nuclei can be treated by configuration interaction (CI) techniques (interacting shell model (green)). For heavy nuclei, the density functional theory, based on self-consistent/mean-field theory (blue), is the tool of choice. The red vertical and horizontal lines show the magic numbers, reflecting regions where nuclei are expected to be more tightly bound and have longer half-lives. The anticipated path of the astrophysical r-process responsible for nucleosynthesis of heavy elements is also shown (purple line). Adapted from Ref. \cite{UNEDF}.}
\label{landscape}
\end{center}
\end{minipage}
\end{figure}

There are approximately 3,000 known nuclei, most of them produced in the laboratory, and an estimated 6,000 nuclei that could in principle still be created in nuclear laboratories and in the Cosmos. Understanding the properties of these nuclei is crucial for a complete nuclear theory, for element formation, for properties of stars, and for present and future energy and defense applications.  Figure \ref{landscape} shows the nuclear landscape as a function of neutron and proton number.
Overlaying the nuclear landscape are the regions applicable for the major theoretical approaches and computational techniques utilized in the UNEDF collaboration:  ab initio, configuration interaction, and density functional theory.  Methods are applicable to a particular mass of nuclei and constrained by the computational resources available.  Furthermore, by investigating the intersections and overlaps of these regions, UNEDF members gain valuable input for establishing a robust theory with high-quality predictive power.

\subsection{Strategy Diagram}
The UNEDF collaboration involves a synthesis of different perspectives to face the challenge of understanding the low-energy nuclear many-body problem within the landscape of high-performance computing.  Accomplishing the scientific goals requires development of integrative solutions that extend beyond the capabilities of a single domain while maintaining a clear, unified strategy.

\begin{figure}[htdp]
\begin{minipage}{6in}
\begin{center}
\includegraphics[bb=0 0 600 800,clip,trim=0cm 0cm 0cm 1cm,width=0.99\textwidth]{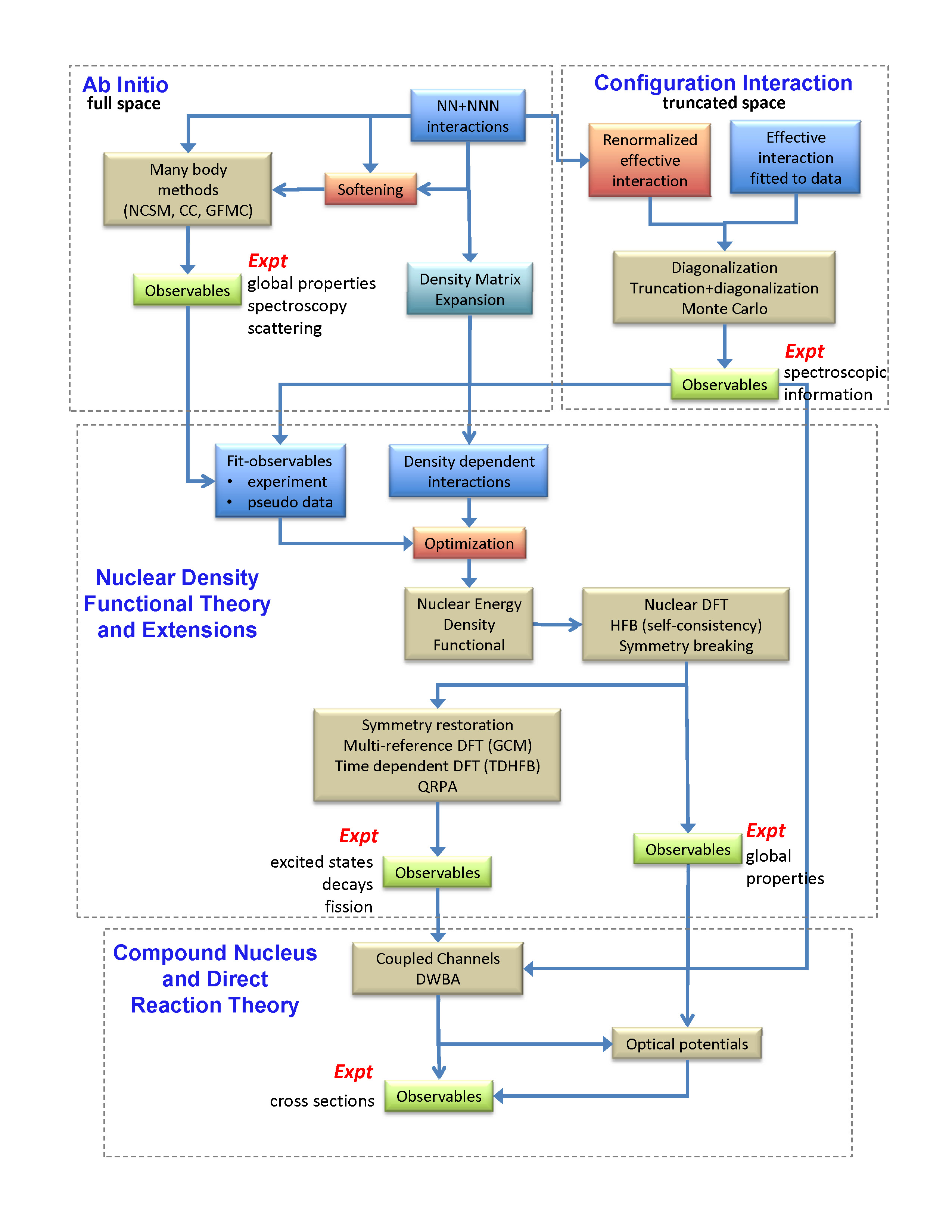}
\caption{UNEDF strategy diagram identifying the interconnections between the four primary focus areas: ab initio, configuration interaction, nuclear density functional theory and extensions, and compound nuclear and direct reaction theory.  Taken from http://unedf.org.}
\label{strategy}
\end{center}
\end{minipage}
\end{figure}

The UNEDF strategy diagram shown in Fig.~\ref{strategy} displays the cohesive view of the collaboration's strategy \cite{NPN}.
This diagram shows the major scientific focus areas and provides necessary granularity to the overlapping computational methods seen in Fig.~\ref{landscape}.  Through the strategy diagram, UNEDF conveys the interdependence of the various focus areas of the collaboration and identifies the challenges where multidomain expertise is necessary. It also provides a meaningful division of labor as well as a global perspective on the impact from individual efforts. Furthermore, the diagram in Fig.~\ref{strategy}  helps to identify and foster unexpected cross-cutting physics research and shared computational challenges. Clearly outlining the efforts serves to maintain focus and heighten the collaborative spirit within the group.

\subsection{Measures of Success}
\begin{wrapfigure}[20]{l}{0.5\textwidth}
\begin{center}
\includegraphics[bb=0 0 350 250,clip,trim=0cm 0cm 0cm 0.5cm,width=0.5\textwidth]{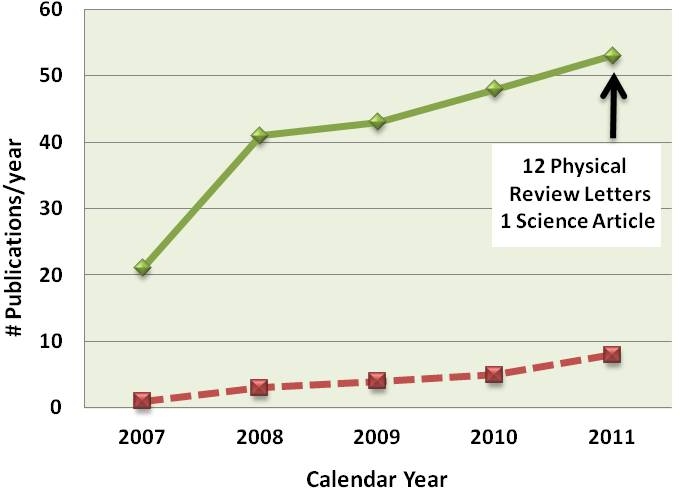}
\caption{Publications produced yearly through the UNEDF collaboration.  The solid line shows all publications and the dashed-line shows cross-domain authorship on publication.  {\it Number of publications for 2011 is incomplete.} }
\label{publications}
\end{center}
\end{wrapfigure}

The success of the SciDAC UNEDF collaboration can be quantitatively measured by the scientific impact of the research performed.  Figure \ref{publications} shows the number of publications resulting from UNEDF research over five years.  Important to note are the high number of {\it Physical Review Letters} as the collaboration has matured; a {\it Science} highlight \cite{BULGAC} 
in 2011; and cross-domain authorship in nuclear physics, physics, and computing publications.

The UNEDF effort has also placed great importance on recruiting and retaining the next generation leaders in low-energy nuclear physics, applied mathematics, computer science, and high-performance computing.  Annually, it has provided training to approximately 30 young researchers, including postdocs and graduate students.  Through experience in the UNEDF collaboration, researchers have received job placement in national laboratories and faculty positions at universities, as well as received various funding and awards.  A list of publications, one-page highlights,  and additional information on awards and appointments can be found at the UNEDF website.

Qualitatively, UNEDF success can be seen in the increased collaboration across domains.  These connections help lay the foundation for future research and address the challenges posed by emerging architectures.

\section{High-Performance Computing Resources}
Access to leadership-class computing resources and large compute time allocations are critical to the scientific investigations of many UNEDF members.  Through the competitive INCITE (``Innovative and Novel Computational Impact on Theory and Experiment'') program, UNEDF members have been awarded large allocations on leadership-class computing resources at the Oak Ridge Leadership Computing Facility (OLCF) and the Argonne Leadership Computing Facility (ALCF). Computing resources include the following:
\begin{description}
  \item[Intrepid (ALCF),]an IBM Blue Gene/P system with 40 racks containing 1024 nodes per rack and 850 MHz quad-core processors and 2 GB RAM per node.  Intrepid currently provides users with 163,840 cores, roughly 82 TB of memory, 7.6 PB of disk space, and 88 GB/s of disk bandwidth.
  \item[Jaguar (OLCF),]a Cray XT with two partitions.  The XT4 partition contains 7,832 compute nodes with quadcore AMD Opteron 1354 (Budapest) processors and 8 GB RAM per node, totaling 31,328 processing cores.  The XT5 partition contains 18,688 compute nodes with dual hex-core AMD Opteron 2435 (Istanbul) processors and 16 GB RAM per node, totaling 224,256 processing cores.  Jaguar currently provides users with a peak performance of 2.332 PF, 299 TB of system memory, 10 PB of disk space, and 240 GB/s of disk bandwidth.  {\it Note:  The XT4 partition was the primary resource in 2008 and was retired in 2011.  The XT5 partition was available in  2009 with dual quad-core AMD Opteron 2356 (Barcelona) processors and was upgraded to hex-core in 2010.}
\end{description}

\begin{figure}[htdp]
\begin{minipage}[b]{3in}
\begin{center}
\includegraphics[bb=0 0 325 225,clip,trim=0cm 0cm 0cm 0cm,width=0.95\textwidth]{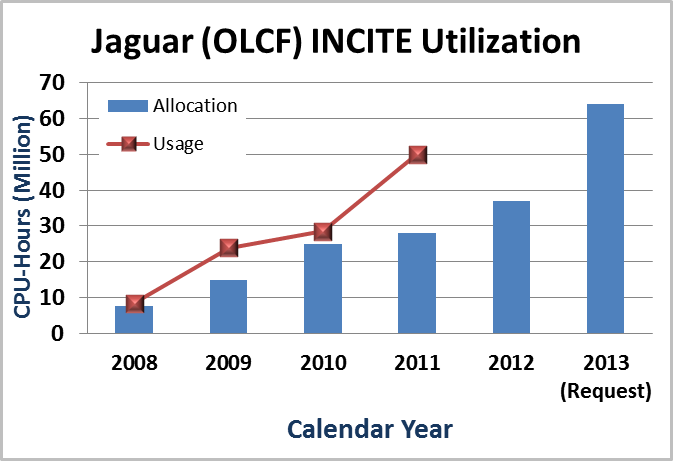}
\caption{INCITE allocation and utilization of the Jaguar supercomputer at the OLCF (CY 2008-2013)}
\label{jaguar}
\end{center}
\end{minipage}
\hspace{1.5pc}
\begin{minipage}[b]{3in}
\begin{center}
\includegraphics[bb=0 0 325 225,clip,trim=0cm 0cm 0cm 0cm,width=0.95\textwidth]{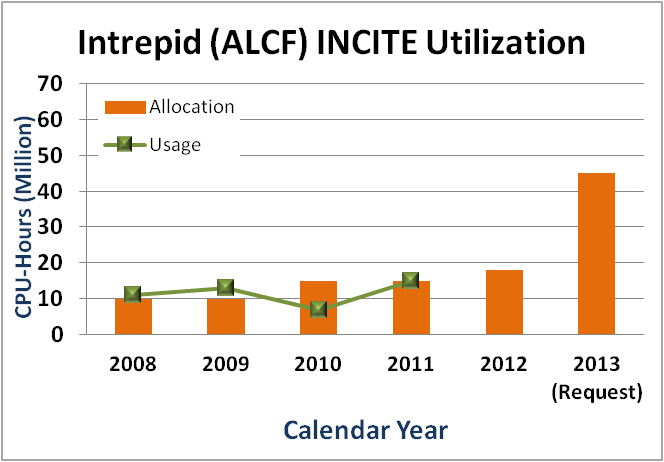}
\caption{INCITE allocation and utilization of the Intrepid supercomputer at the ALCF (CY 2008-2013)}
\label{intrepid}
\end{center}
\end{minipage}
\end{figure}

Through the UNEDF collaboration, members have been able to continuously scale codes to efficiently utilize these ever-increasing resources.  Figures \ref{jaguar} and \ref{intrepid} show the INCITE allocations awarded to UNEDF collaborators and the CPU-hour utilization starting from 2008.  These figures highlight the increasing demand for computing time in low-energy nuclear physics research.  The combined utilization across Jaguar and Intrepid in 2008 was nearly 20 million CPU-hours and has increased more than threefold in 2011.  For the 2012 calendar year, UNEDF members were granted the sixth largest allocation of the 60 INCITE projects awarded.  These statistics show that low-energy nuclear physics research is dependent on high-performance computing and equally that low-energy nuclear physics is a scientific driver in HPC.

Leadership-class utilization is an important metric for assessing the need for and usability of capability computing resources.  Capability computing is defined as ``using the maximum computing power to solve a large problem in the shortest amount of time,'' thus solving ``a problem of a size or complexity that no other computer can'' \cite{Graham}. This is in contrast with capacity computing, which uses ``efficient cost-effective computing power to solve somewhat large problems or many small problems.''  At the OLCF, utilization of Jaguar is binned into three job-size categories: usage of less than 20\%, between 20 and 60\%, and greater than 60\% of the computing resource for a single job. The typical scale of a code appropriate for a leadership-class system is utilization of greater than 20\% of the resource for a single calculation. Job sizes less than 20\% of the resource can typically fit onto smaller capacity systems.
\begin{table}[hb]
\caption{\label{leadership}Utilization of Jaguar Supercomputer}
\begin{center}
\begin{tabular}{lrcc}
\br
Year& Allocation &Usage (CPU-Hours) & Leadership\\
\mr
2011&28,000,000&50,076,810&66\%\\
2010&25,000,000&28,465,982&61\%\\
2009&15,000,000&23,859,172&78\%\\
2008&7,500,000&8,432,335&65\%\\
\br
\end{tabular}
\end{center}
\end{table}

\begin{wrapfigure}[16]{l}{0.65\textwidth}
\begin{center}
\includegraphics[bb=0 0 475 275,clip,trim=0cm 0cm 0cm 1cm,width=0.65\textwidth]{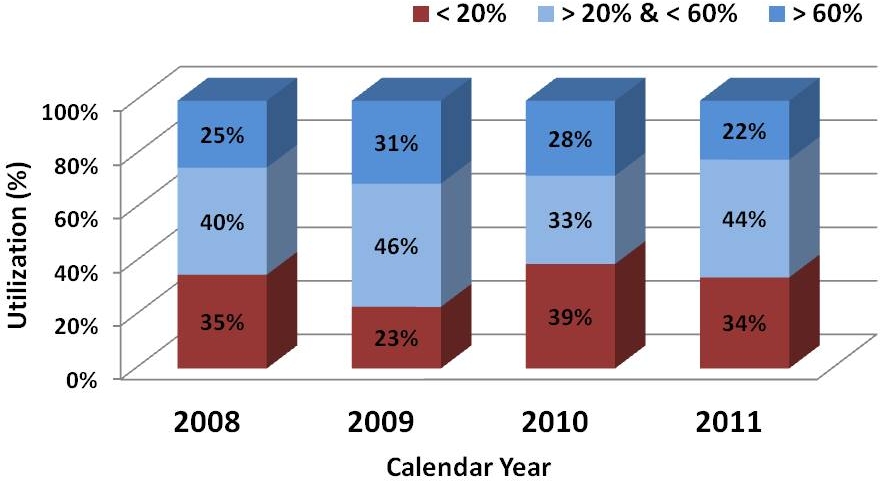}
\caption{Utilization of Jaguar.}\label{utilization}
\end{center}
\end{wrapfigure}

Table \ref{leadership} and Fig.~\ref{utilization} show utilization by UNEDF projects of the Jaguar supercomputer binned by job size.  Table \ref{leadership} shows that UNEDF projects consistently use over 60\% of their allocation for leadership-size jobs.  It is important to note that in 2008, when Jaguar was an XT4, leadership-class jobs used more than 6,266 cores, in 2009 for the Jaguar XT5 quad-core jobs used more than 29,901 cores, and since 2010 for the Jaguar XT5 hex-core jobs used more than 44,852 cores. Figure \ref{utilization} provides additional granularity to show that UNEDF projects require 60\% of the resource for a single computational run for nearly 25\% of their usage.  This shows the success of UNEDF collaborations to continually meet the changing architecture and growing size of computing systems.

\begin{figure}[htb]
\begin{minipage}{6in}
\vspace{.5cm}
\begin{center}
\includegraphics[bb=0 0 450 350,clip,trim=0cm 0cm 1.5cm 0cm,width=0.75\textwidth]{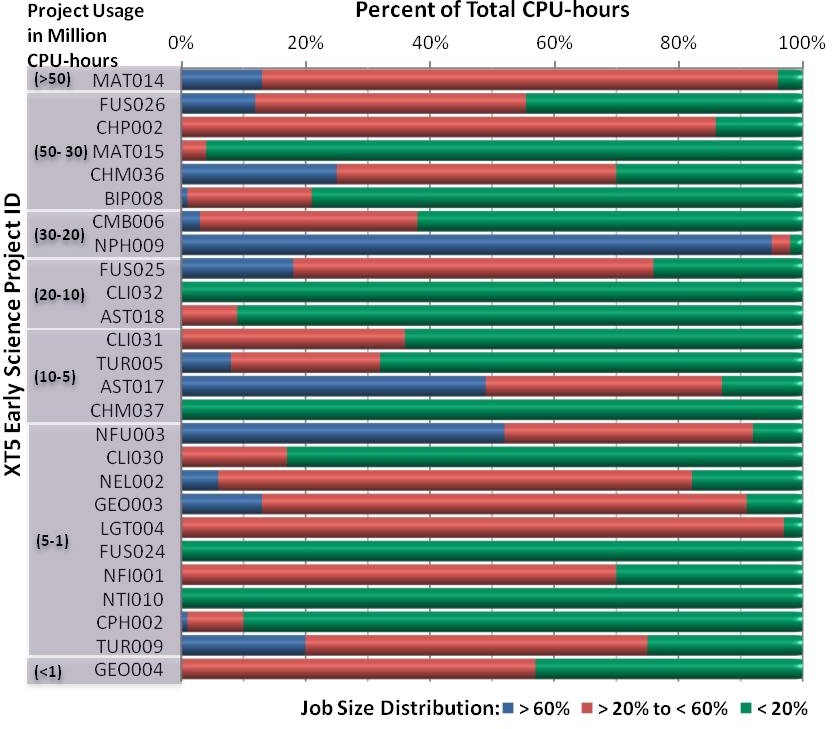}
\caption{Utilization of Jaguar XT5 by job size for Early Science projects \cite{exascale}.}\label{earlyscience}
\end{center}
\end{minipage}
\end{figure}

Additional computing time was provided in 2009 through the OLCF Early Science period prior to transitioning the general user population onto the Jaguar XT5 quad-core.  At that time, the XT5 partition had 18,688 compute nodes with dual quadcore AMD Opteron 2356 (Barcelona) processors, totaling 149,504 processing cores.  The XT5 partition became available to the larger user community in July 2009; for the first half of 2009, during its transition-to-operations period, it was open only to select Early Science users. UNEDF members were awarded 30 million CPU-hours for an Early Science project on the XT5.  Figure \ref{earlyscience} shows the utilization by job size of over 350 million CPU-hours over six months by 26 projects.  The low-energy nuclear physics project labeled NPH009 shows that over 95\% of its utilization was at leadership class \cite{exascale}.

Close collaboration between nuclear physicists, applied mathematicians and computer scientists enable UNEDF research to effectively utilize high-performance computing resources, leading to the science highlights presented here.

\section{High-Performance Computing Enhances \textit{Ab Initio} Nuclear Structure Calculations}
\begin{figure}[h]
\begin{minipage}[b]{3in}
\begin{center}
\includegraphics[bb=0 0 2400 1800,clip,trim=0cm 0cm 0cm 0cm,width=0.95\textwidth]{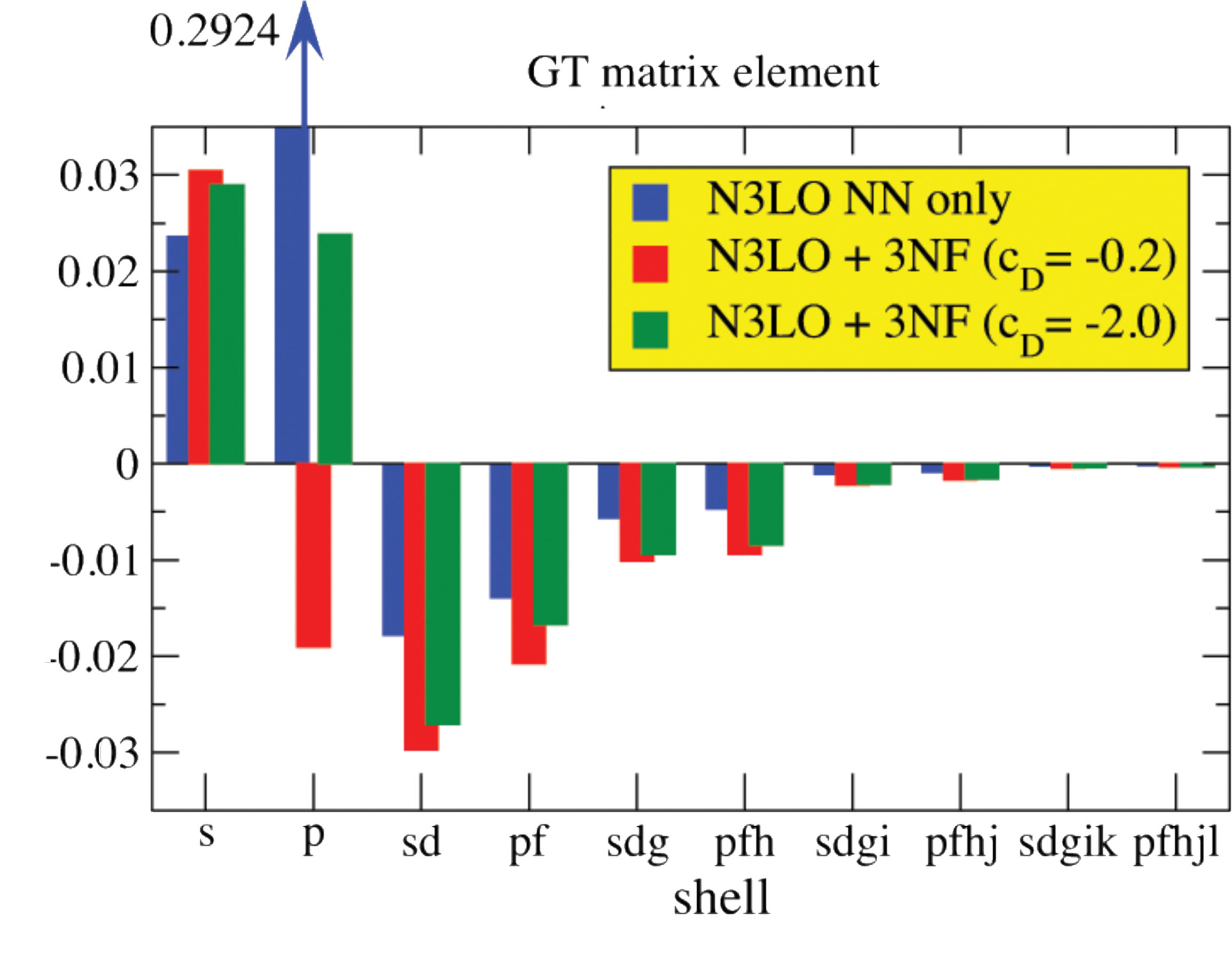}
\caption{MFDn simulation shows three-body forces are necessary to explain the anomalously long half-life of isotope carbon-14 used in carbon dating \cite{MFDn}.}
\label{fig1}
\end{center}
\end{minipage}
\hspace{1.5pc}
\begin{minipage}[b]{3in}
\begin{center}
\includegraphics[bb=0 0 450 350,clip,trim=0cm 0cm 0cm 0cm,width=0.95\textwidth]{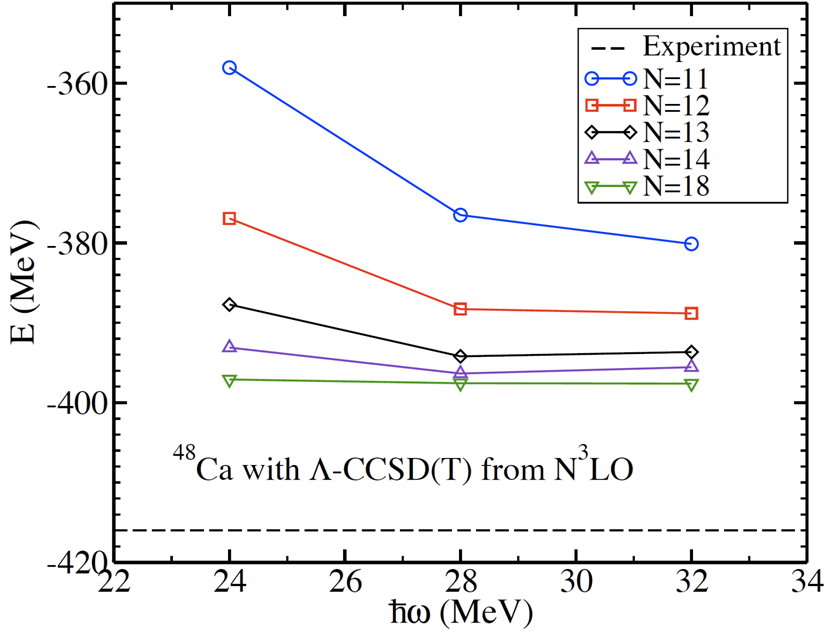}
\caption{NUCCOR calculations for medium-mass nuclei calcium-48 shows three-nucleon forces account for the missing binding energy \cite{HAGEN}.}
\label{fig2}
\end{center}
\end{minipage}
\end{figure}

\textit{Ab initio}, or from first principles, nuclear structure calculations have made major advances under UNEDF toward effectively utilizing high-performance computing resources and transforming to meet the challenges posed by emerging architectures.  Ab initio techniques provide a fine-grained method for studying nuclei and the nuclear interaction, but they often come with a high computational cost.  They are necessary to the UNEDF effort by providing ``control data'' to constrain more general functionals and test candidate energy density functionals even for systems not experimentally accessible.  UNEDF collaborators continue to scale ab initio nuclear structure simulations and perform the largest and most accurate calculations currently possible on both Jaguar (ORNL) and Intrepid (ANL).

Examples of UNEDF-directed advances include development of the Asynchronous Dynamic Load Balancing (ADLB) software library by using {\it Green's function Monte Carlo} (GFMC) calculations as a testbed. The ADLB library has enabled GFMC to run efficiently on over 100,000 cores on Intrepid \cite{LUSK}.  Utilizing Jaguar, UNEDF applications \textit{Nuclear Coupled-Cluster - Oak Ridge} (NUCCOR) and \textit{Many Fermion Dynamics-nuclear} (MFDn) have undergone considerable code and algorithm development.  Improvements include implementation of a hybrid MPI and OpenMP approach for efficient memory management, memory-aware algorithms, and integration of libraries and tools to enable further scaling for higher-precision calculations.

Recent scientific breakthroughs include using NUCCOR to calculate medium-mass nuclei from the ground up starting from nucleon degrees of freedom, such as $^{48}$Ca shown in Figure \ref{fig2} \cite{HAGEN}. Results show chiral nucleon-nucleon interactions perform remarkably well; the 400\,keV per nucleon missing binding energy in $^{48}$Ca can be attributed to chiral three-nucleon forces missing in calculations \cite{HAGEN}.  Another recent highlight explains the useful but anomalously long lifetime of $^{14}$C by identifying the critical role of the three-nucleon force in its beta decay seen in Fig.~\ref{fig1} \cite{MFDn}.  These calculations involve diagonalization of a Hamiltonian matrix of dimension 2 billion, using 214,668 cores on the Jaguar supercomputer at ORNL under the Early Science projects allocation of 30 million CPU-hours.

\section{Advanced Algorithms and Tools Define New Generation Energy Density Functionals}
\begin{figure}
\begin{minipage}{6in}
\begin{center}
\includegraphics[bb=0 0 4000 1800,clip,trim=0cm 0cm 0cm 0cm,width=0.95\textwidth]{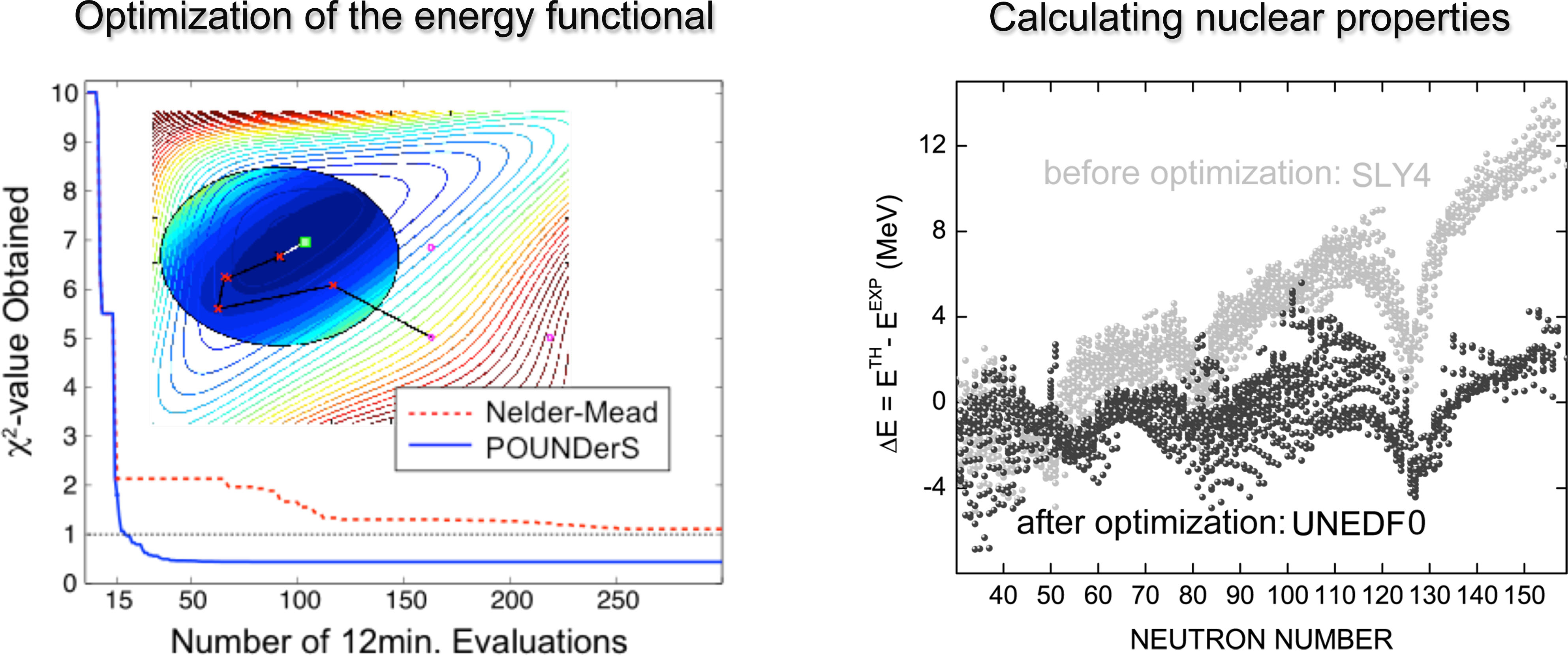}
\caption{The optimization algorithm POUNDerS yields dramatic computational savings over alternative optimization methods (left).  The resulting parameterization UNEDF0 obtained by using the POUNDerS algorithm provides a baseline of nuclear ground-state properties to compare with future functionals (right) \cite{Pounders}.}\label{fig3}
\end{center}
\end{minipage}
\end{figure}

The UNEDF project has devoted considerable effort to develop and improve the algorithmic and computational infrastructure needed to optimize candidate energy density functionals (EDF).  These developments have resulted in new optimization tools that are broadly available to other science domains.  An example is the optimization algorithm POUNDerS, which not
only provides a computational savings over other methods, as shown in Fig.~\ref{fig3}, but greatly improves the time to solution to test candidate EDFs. With the derivative-free POUNDerS algorithm, the resulting parameterization of existing data yielded UNEDF0, which sets a solid baseline of nuclear ground-state properties to compare\begin{figure}[hb]
\begin{minipage}{6in}
\begin{center}
\includegraphics[bb=0 0 4300 1900,clip,trim=0cm 0cm 0cm 0cm,width=0.95\textwidth]{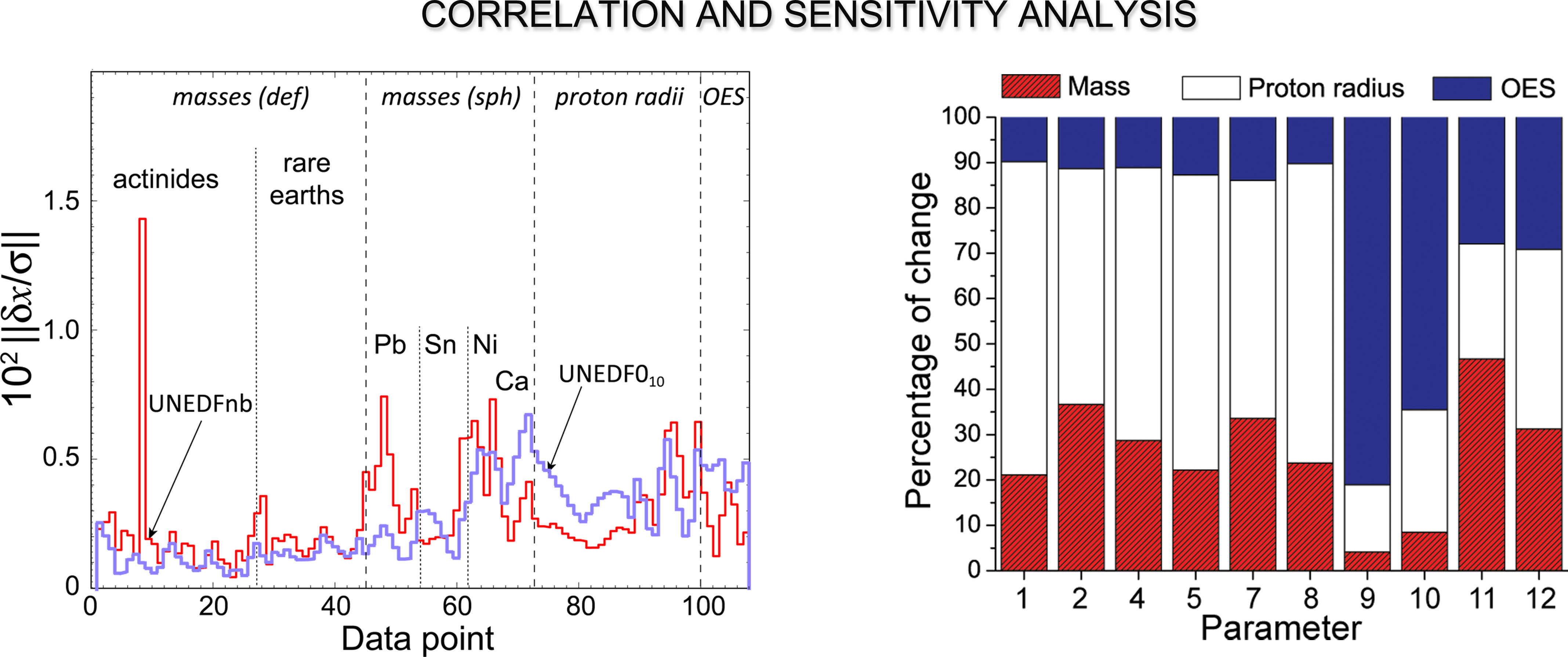}
\caption{Statistical tools are used to deliver uncertainty quantification and error analysis for theoretical studies as well as to assess new experimental data \cite{Pounders}.}\label{fig4}
\end{center}
\end{minipage}
\end{figure} with future functionals shown in Figure \ref{fig3} \cite{Pounders}. Continuing work involves utilizing this approach to study new hybrid functionals with microscopic input from chiral effective field theory \cite{Stoitsov}.

These new tools enable for the first time, a consistent method for uncertainty quantification and correlation analysis to estimate errors and significance as a first step toward a formal process for future verification and validation.   Included in the UNEDF project are the development and application of statistical tools, particularly important for directing future experiments by providing analysis of the significance of new experimental data. For example, the sensitivity of two optimized functionals to particular data is shown in Fig.~\ref{fig4} \cite{Pounders}. Such capabilities have not been previously available in the low-energy nuclear theory community but are increasingly important as new theories and computational tools are applied to new nuclear systems and to conditions inaccessible to experiment.

\section{Massively Parallel Algorithms Open Cold Atoms as a Testing Ground}

UNEDF theorists have made important contributions to the study of strongly coupled superfluid systems such as ultracold Fermi atoms, which show many similarities to the cold nuclear matter found in the crust of neutron stars. Cold atoms make excellent laboratories for testing and \begin{wrapfigure}[50]{l}{0.5\textwidth}
\begin{minipage}{3in}
\begin{center}
\includegraphics[bb=0 0 1000 680,clip,trim=0cm 0cm 0cm 0cm,width=0.95\textwidth]{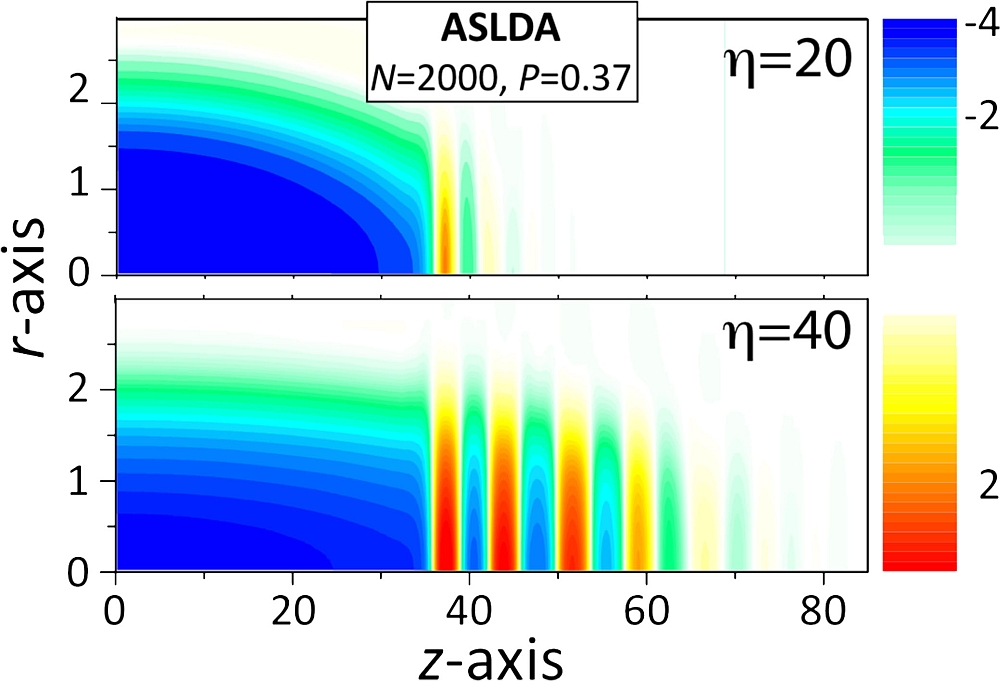}
\caption{ASLDA simulation with strongly interacting spin-imbalanced atomic gases in extremely elongated traps \cite{PEI}.}\label{fig5}
\end{center}
\end{minipage}
\begin{minipage}{3in}
\begin{center}
\includegraphics[bb=0 0 1000 450,clip,trim=0cm 0cm 0cm 0cm,width=0.95\textwidth]{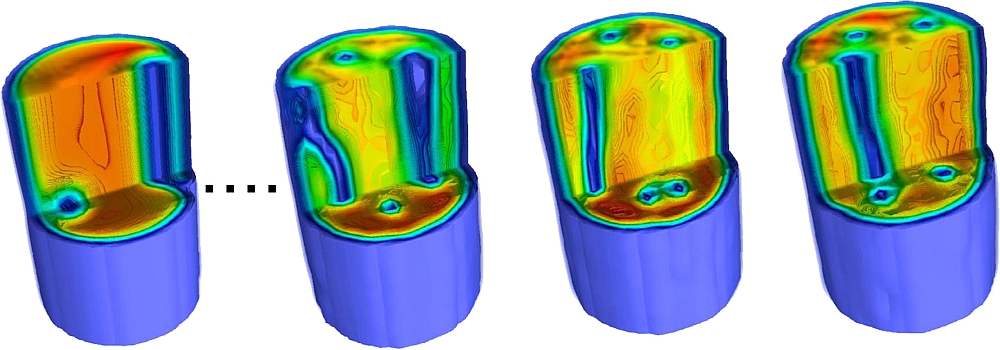}
\caption{TDSLDA simulation shows the ball-and-rod excitation of a unitary fermi gas with vortex formation \cite{BULGAC}.}\label{fig6}
\end{center}
\end{minipage}
\end{wrapfigure}improving the computational methods to be used for nuclei.  Cold-atom systems also allow for predictions of superfluid DFT that are testable against experiment.

UNEDF developments using cold atoms as a testing ground include adding new algorithms to existing applications, such as adapting the antisymmetric superfluid local density approximation (ASLDA) to an existing massively parallel nuclear DFT code with strongly interacting spin-imbalanced atomic gases in extremely elongated traps, seen in Fig.~\ref{fig5} \cite{PEI}.  Another major UNEDF development is implementation of the time-dependent superfluid local density approximation (TDSLDA) on a 3D spatial lattice \cite{BULGAC}. Unlike previous methods, the UNEDF implementation eliminates the need for matrix operations, allowing it to accommodate a basis set that is 2-3 orders of magnitude larger than other approaches.  Calculations \cite{BULGAC} were performed by using 97\% of Jaguar to simulate the unitary gas (e.g., vortex formation) and a heavy nucleus under the action of various external fields, seen in Fig.~\ref{fig6}. While still exploratory, these first-time simulations of this kind for fermion superfluids serve as proof of principle for an eventual treatment of neutron-induced fission.

\section{HPC Empowers New Era for Nuclear Reaction Theory}

One of the principal aims of the UNEDF project is to calculate nucleon-nucleus reactions crucial for \begin{wrapfigure}[26]{l}{0.5\textwidth}
\begin{center}
\includegraphics[bb=0 0 700 500,clip,trim=0cm 0cm 0cm 0cm,width=0.5\textwidth]{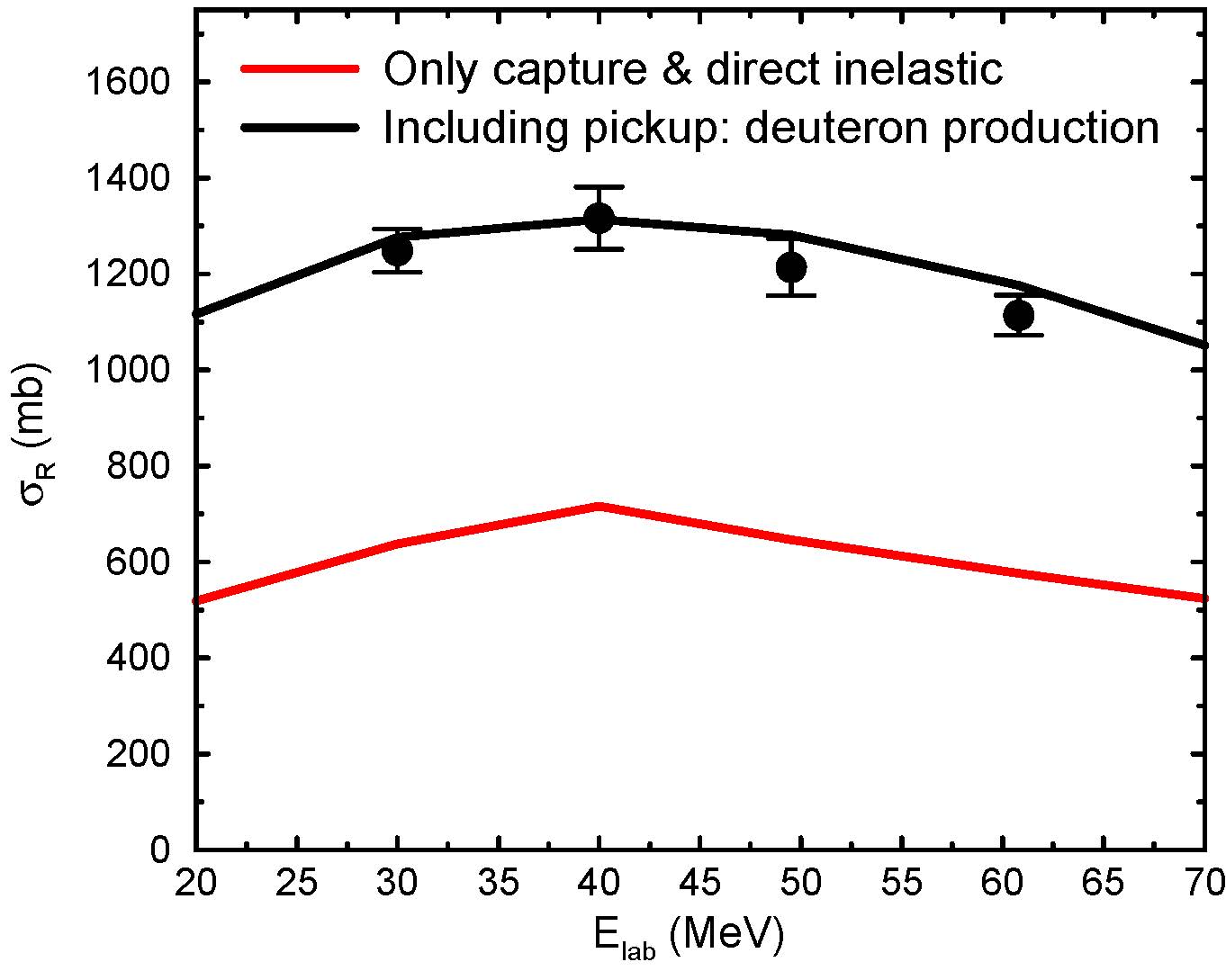}
\caption{New methods for calculating reaction cross-sections (black) show good agreement with experimental data.  Total reaction cross section as a function of the incident energy for the reaction p + 90Zr using the Gogny D1S force. The results are shown for couplings to the inelastic RPA states (red line), and to the inelastic and transfer channels with nonorthogonality corrections (black line) \cite{IAN}.}\label{ian}
\end{center}
\end{wrapfigure}astrophysics, nuclear energy, radiobiology, and national security, for which extensions of standard phenomenology is insufficient.  Under the UNEDF effort, neutron reactions on heavier nuclei are being modeled by using DFT results to predict not just bound states but also scattering states for nucleons. As shown in Fig.~ś\ref{ian}, the calculated reaction cross-sections agree well with experimental data.  For the first time, a complete microscopic calculation using basic interactions between nucleons can be used to predict reaction observables with low-incident energy \cite{IAN}.  This technology provides the basis for future calculations of unstable species outside the range of experiment.

Another important capability for reactions is the calculation of level densities, which provides insight to the interactions inside the system. A new proton-neutron algorithm for the parallel JMoments code was recently designed and implemented, which scales to tens of thousands of cores and greatly increases the code's overall performance. This development opens the door to calculating accurate, nuclear level densities and reaction rates for a large class of nuclei \cite{IAN2}.

\section{Outlook}
\label{sec6}
The UNEDF collaboration has provided fertile ground for new and continuing growth between applied mathematics, computer science, and nuclear physics.  Over the past five years, the collaboration has established cross-disciplinary working relationships to facilitate future efforts and has matured to adequately address new challenges in verification and validation, workflow, visualization, and new programming models with changing architectures.  Reaching next-generation science objectives requires computational resources several orders of magnitude beyond what is currently available.  Adapting to these changes will take conscious planning and purposeful action.  The members of the collaboration are well positioned to meet these disruptive changes through the close working relationship established through UNEDF.  UNEDF, and similar future collaborations, will continue to develop key computational codes and algorithms for reaching the goal of solving the nuclear quantum many-body problem, thus paving the road to the comprehensive model of the atomic nucleus.

\section{Acknowledgments}

The UNEDF SciDAC collaboration is supported by the U.S. Department of
Energy (DOE) under grant No. DOE-FC02-09ER41583. This work was also supported
by DOE Contract Nos.~DE-FG02-96ER40963 (University of Tennessee), DE-AC05-00OR22725 with UT-Battelle, LLC (Oak Ridge National Laboratory), DE-FG05-87ER40361 (Joint Institute for Heavy Ion Research), DE-AC02-06CH11357 (Argonne National Laboratory), DEFC02-09ER41582 (Iowa State University), and DE-FG02-87ER40371(Iowa State University). Computational resources were provided through an INCITE award ``Computational
Nuclear Structure'' by the National Center for Computational Sciences (NCCS) and
National Institute for Computational Sciences (NICS) at Oak Ridge National
Laboratory.


\section*{References}

\begin{thebibliography}{13}
\bibitem{UNEDF} Bertsch G F, Dean D J and  Nazarewicz W 2007 {\it SciDAC Review} {\bf 6} 42
\bibitem{NPN} Furnstahl R J (for the UNEDF Council) 2011 Nuclear Physics News {\bf 21} 18
\bibitem{exascale} Joubert W, Kothe D and Nam H 2009 {\it Preparing for Exascale: ORNL Leadership Computing Facility Application Requirements and Strategy} (Oak Ridge: ORNL)
\bibitem{Graham}Graham S L, Snir M, and Patterson C A 2005 {\it Getting Up to Speed: The Future of Supercomputing} (National Academies Press: Washington, D.C.)
\bibitem{BULGAC}Bulgac A et al 2011 {\it Science} {\bf 332} 1288
\bibitem{LUSK} Lusk E et al 2010 {\it SciDAC Review} {\bf 17} 30
\bibitem{PEI} Pei J C et al 2010 {\it Phys. Rev.} A {\bf 82} 021603(R)
\bibitem{HAGEN} Hagen G et al 2010 {\it Phys. Rev.} C {\bf 82} 034330
\bibitem{MFDn} Maris P et al 2011 {\it Phys. Rev. Lett.} {\bf 106} 202502
\bibitem{Pounders}Kortelainen M et al 2010 {\it Phys. Rev.} C {\bf 82} 024313
\bibitem{Stoitsov}Stoitsov M et al 2010 {\it Phys. Rev.} C {\bf 82} 054307
\bibitem{IAN} Nobre G et al 2011 {\it Phys. Rev.} C {\bf 84} 064609
\bibitem{IAN2}Senkov R and Horoi M 2010 {\it Phys. Rev.} C {\bf 82} 024304
\end{thebibliography}
\end{document}